# The $\chi^2$-test, the Muon AMM and Karl R. Popper

**Giuseppe Iurato**
*Department of Physics, University of Palermo, IT*
*Department of Mathematics and Informatics, University of Palermo, IT*

E-mail: giuseppe.iurato@unipa.it

**Abstract.** In this very brief note, we only wish to identify a simple but notable epistemological basis, concerning the Karl R. Popper philosophy of science thought, into the realm of the experimental proves of Fundamental Physics.

## 1. Introduction

One of the central problems in the theory of data elaboration of experimental physics is that of determining the probability distributions of the various measures obtained from an experimental measurement which, in general, belongs to the class formed by the Gauss, Bernoulli and Poisson distributions. The tests for the deviation of the observed distribution of experimental data from a presumed theoretical distributions, provide criteria for deciding with what approximation the former is in accordance with the latter.

There exist specific tests for each distribution (as, for instance, the $\beta$-skewness and the $\beta_2$-flatness for the Gauss distribution), as well as some related to a general theoretical distribution, like the well-known (among the non-parametric[1] statistical tests for goodness fit) $\chi^2$-test and the Kolmogorov one, which turn out to be independent from the presumed distributions (distribution-free).

In what follows, we recall the main outlines on the $\chi^2$-test, introduced by K. Pearson at the very beginning of the 1900. We mainly follow the good exposition given by (Taylor, 1986).

## 2. The quantitative $\chi^2$-test: brief outlines

Let $X$ be a random variable defined in $[a, b] \subseteq \mathbb{R}$ and $\mathcal{F}$ a partition of it into $n$ subintervals, and let $O_k$ and $E_k$ be respectively the observed and the expected values of $X$ which fall into the $k$-th subinterval. The hypothesis (often denoted with $H_0$) according to which the observed values follow a given preassigned theoretical probability distribution which predict the given expected values, clearly depends by the various related deviations $O_k - E_k$, so that it turns out to be natural considering the following number

(1) $$\chi^2 = \sum_{k=1}^{n} \frac{(O_k - E_k)^2}{E_k}$$

said to be *chi-squared*, which is a good estimate for the accordance between the observed distribution and the presumed one. Following (Montanari & Poppi, 1982), this test is significative at least when $n \geq 40$ and allows to decide whether the deviations between the experimental and hypothesized theoretical distributions (*hypothesis $H_0$*) are due to the casualness or not.

Since it is expected that each term of this sum be about 1, then we would have approximately $\chi^2 \leq n$ if there is a good accordance between the observed distribution and the hypothesized one, otherwise we would have $\chi^2 \gg n$. A better accordance estimate is carried out if one compares the chi-squared with the statistical freedom degrees $d = n - c$ instead of the simple interval number $n$, where $c$ is the number of parameters and/or relations related to the experimental data, which is also called *constrained* number[2]. Therefore, it is possible to prove as the expected value of $\chi^2$ is just $d$, so that if $\chi^2 \gg d$ then it is very likely that the hypothesized distribution does not agree with the experimental one. Often it is considered the so-called *reduced chi-squared* defined as $\bar{\chi}^2 = \chi^2/d$, so that if one obtains an estimate of the $\bar{\chi}^2$ less or equal to 1, then the initial hypothesis is valid, whereas if $\bar{\chi}^2$ is much greater than 1 it isn't.

---

[1] Albeit such a term is quite incorrect; in this regards, see (Girone & Salvemini, 2000, § 25.2).

Now, the quantitative $\chi^2$-test tries to determine how a reduced chi-squared greater than one may be considered invalidating of our hypothesis $H_0$. To answer to such a question, one considers the probability[3] $P(\bar{\chi}^2 \geq \bar{\chi}_0^2)$ of obtaining a reduced chi-squared greater or equal to that experimentally computed: if this is quite high, then our value[4] $\bar{\chi}_0^2$ is acceptable because is very likely to obtain an higher value of it, so that the latter is one of the possible lower values. Hence, it is important at first to determine at what lower value of the probability[5] $P(\bar{\chi}^2 \geq \bar{\chi}_0^2)$ the given hypothesis $H_0$ is refusable. In general, it is chosen two (but not the only ones possible) rejection confidence levels $P_{sign}$ under to which there is disagreement, that is to say when $P(\bar{\chi}^2 \geq \bar{\chi}_0^2) < P_{sign}$: the $P_{sign} = 5\%$ level and the $P_{sign} = 1\%$ level.

In practice, chosen the confidence level $P_{sign} = \int_{\chi_0^2}^{+\infty} f(\chi^2) d(\chi^2)$ once known the value $\chi_0^2$ computed through (1), then it is possible to determine a value of $\chi^2$ such that[6] $P(\chi^2 \geq \chi_0^2) = P_{sign}$, the left hand side of this equation being computed through proper probability tables in dependence on $d$ and on the $\chi^2$ probability distribution $f(\chi^2)\, d(\chi^2) = 2^{-\frac{d}{2}} \left(\Gamma\left(\frac{d}{2}\right)\right)^{-1} (\chi^2)^{\frac{d-2}{2}} e^{-\frac{\chi^2}{2}} d(\chi^2)$. Thus, if it results to be $\chi^2 < \chi_0^2$, then $H_0$ is rejected at the given confidence level $P_{sign}$, whereas, if it turns out to be $\chi^2 > \chi_0^2$, then $H_0$ is considered to be valid at the confidence level $P_{sign}$, at least until further information[7] (see (Porto, 1987, Capitolo VIII, § 3)). Indeed, the $\chi^2$-test is definitive only and only when it leads to a rejection of the assumed hypotheses, as, for instance, the Muon Anomalous Magnetic Moment (AMM) determination story shows (see (Jegerlehner, 2008), (Melnikov & Vainshtein, 2006) and references therein), as regards the fundamental experimental ascertainment of the Dirac QED hypothesis $g = 2$, which is one of the main tests provided by the Standard Model. In fact, since 1950s (with the first pioneering Antonino Zichichi's research group works at CERN) till lately, this fundamental hypothesis hasn't been rejected but ever more confirmed with increasingly improvements and better precision, just through the $\chi^2$-test.

In this last sense, such a $\chi^2$-test, with its consequences, might also be considered as a powerful formal tool which gives rise to possible links with the main points of the well-known Karl R. Popper *falsificationism*, in this case applied to Fundamental Physics.

---

[2] Roughly speaking, $d$ provides the number of independent random variables.
[3] Which is computed respect to the same theoretical probability distribution already considered.
[4] With $\chi_0^2$ (or $\chi_{exp}^2$), we denote the observed (or experimental) value of the chi-squared computed by means of (1), whereas with $\chi_{Th}^2$, or simply with $\chi^2$, we denote the theoretical one; likewise for $\bar{\chi}^2$.
[5] The value $P(\bar{\chi}^2 \geq \bar{\chi}_0^2)$ is given by the so-called *Helmert-Pearson $\chi^2$ probability distribution*; see, for instance, (Stoka, 1991, Capitolo VIII, § 8.1).
[6] This simply means that $P_{sign}$ is the probability with which $\chi^2$ may overcomes $\chi_0^2$.
[7] Moreover, in this last case, if the related $P(\chi^2 \geq \chi_0^2)$ is also low, then this suggests the presence of systematic errors, so that further experimental improvements are required.